\documentclass[onecolumn,showpacs,preprintnumbers]{revtex4}
\usepackage[english]{babel}
\usepackage{graphicx}
\usepackage{dcolumn}
\usepackage{ifthen}
\usepackage{bm}

\begin{document}

\title{The self-consistent microscopic model of an energy spectrum of a superfluity $^{4}$He with the
Hermitian form of Bogoliubov-Zubarev Hamiltonian.}
\author{K.V. Grigorishin}
\email{konst_dark@mail.ru}
\author{B.I. Lev}
\email{bohdan.lev@gmail.com} \affiliation{M.M.Bogolyubov Institute
for Theoretical Physics of the Ukrainian National Academy of
Sciences, 14-b Metrolohichna str. Kiev, 03680, Ukraine.}
\date{\today}

\begin{abstract}
Based on the collective variables representation with the
Hermitian form of Bogoliubov-Zubarev Hamiltonian the
self-consistent oscillator model of a ground state and excited
states of Bose liquid has been proposed. The new method of
calculation of anharmonic terms in this Hamiltonian and its
interpretation have been presented. The dispersion equation for a
collective excitation in a superfluid $^{4}$He has been obtained
in self-consistent way, where real and virtual processes of decay
of a collective excitation were considered. The end point,
determined by a threshold of collective excitation's decay on two
rotons, of the dispersion curve has been obtained and it was shown
that the dispersion curve strongly depends on property of its
stability. An approach with a structure factor has been realized
without using of any adaption parameters. Based on the oscillator
model the new method of self-consistent calculation of a ground
state energy and density of Bose condensate has been proposed. The
model of suppression of Bose condensate has been presented.
\end{abstract}
 \maketitle

\section{Introduction}

Up to present time many various models have been proposed for
microscopic description of a superfluid helium. Most of the
models are based on two approaches. The first one is the theoretical-field formalism.
Its characteristic feature is explicit application of
Bose-condensate (BC) and introduction of the higher BC in some
models \cite{pash1,pash2,pash3,shev}. The second one is the
quantum-mechanical approach, where the Schrodinger equation is
solved for a ground state and for lowest excited states of $N$
interacting particles. This approach originates from the papers of
Feynman and Cohen \cite{feyn} where connection between the
structure factor and the dispersion curve of a collective
excitation (CE) was obtained, however this connection is correct at
small values of a wave vector only ($k\rightarrow 0$). The
quantum-mechanical approach does not use BC explicitly unlike the
theoretical-field approach but it calculates BC knowing the wave
function of a ground state.

The quantum-mechanical approach has been developed essentially due to
use of Bogoliubov-Zubarev formalism \cite{BOGOL2}. In this
representation Hamiltonian of a bosons' system is written in
the terms $\rho_{\textbf{k}}$ and
$\partial/\partial\rho_{\textbf{k}}$ where $\rho_{\textbf{k}}$ is
a Fourier transform of fluctuation of density. As it has been
shown in the papers \cite{vacar1,vacar2} normal motion and
superfluity motion can be separated in the $N$-particle
Schrodinger equation. The oscillations of Bose liquid is obtained
in the harmonic approximation. Zero-point oscillations correspond
to a ground state of liquid, and excited states - to any
collective excitations: phonons, rotons, maxsons and so on.
However the non-harmonic terms of the Hamiltonian play a essential
role. The first correction to the harmonic approximation has been
calculated in above-mentioned papers, the second correction has
been calculated in the paper \cite{tom}. Unfortunately any small
parameters aren't in this expansion and calculation of each new
correction is accompanied by large mathematical difficulties.
Contribution of the anharmonic terms in energy of a liquid has
been evaluated by Brillouin-Wigner perturbation procedure in the
papers \cite{sun,lee}, the method of Green function has been used
\cite{grest}, the "shadow wave function" approach has been
developed \cite{viti}.

The dispersion curve of CE obtained by some authors
\cite{feyn,sun,lee,tom} has satisfactory coincidence with the
experimental spectrum if value of the wave vector is
$k<2.5A^{-1}$. In the region $k>3A^{-1}$ the experimental spectrum
reacher the "shelf", that is hybridization of CE with the
two-roton level is observed \cite{zaw}. The region $k>3A^{-1}$ has
been investigated by the theoretical-field method in \cite{pit},
where it has been shown that the "shelf" on the dispersion curve
caused by decay of CE into two rotons and by the end of
the spectrum. Calculation of the next corrections to the spectrum of CE doesn't give the
end point of a dispersion curve. We have to note that the
potential of interaction between atoms is known badly on
small distances $r<2.5A$. A ground state energy is very
sensitive to interaction on small distances just. The situation is
complicated by fact that He II is liquid however complete theory
of liquid doesn't exist till now. That's why authors proceed from gas representations or the
models of quantum crystal \cite{O'reil,andr,fom}.

Bogoliubov-Zubarev Hamiltonian is
non-Hermitian because the transition from Cartesian coordinates to
the collective coordinates is non-unitary. This Hamiltonian has
been used in the many aforesaid paper
\cite{BOGOL2,grest,stra,tom,vacar1,vacar2,vacar4,vacar5,vacar6}.
But the non-hermicity must lead to violation of some theorems of
quantum mechanics. However an operator of speed $g_{\textbf{k}}$
canonical conjugated with $\rho_{\textbf{k}}$ has been introduced
in the paper of Sunakawa \cite{sun}, and his Hamiltonian
represented by these terms is Hermitian. The Hermitian form of
Bogoliubov-Zubarev Hamiltonian and the Jacobian of transition to
collective coordinates have been obtained in the book
\cite{vacar3}. However the
anharmonic terms of this Hamiltonian has not been calculated.

In the Section \ref{Space} superfluid motion and normal motion
have been separated in the Schrodinger equation with the Hermitian
form of Bogoliubov-Zubarev Hamiltonian. In the Section \ref{RPA}
the oscillator model of Bose liquid has been formulated in the
Random Phase Approximation (RPA). In the Section \ref{Spectrum}
contribution of anharmonic terms of the
Hamiltonian in the spectrum of CE has been calculated.
It has been shown that this correction describes
decay of CE and the dispersion curve has the end point
$k_{\textrm{C}}$. In the Section \ref{Vacuum} a ground
state energy $E_{0}$ has been calculated taking into account anharmonic
terms in the Hamiltonian with help the
oscillator model. The mechanism of suppression of BC
has been described and density of BC at zero
temperature has been calculated.

\section{The equation of motion and the space of collective variables}\label{Space}

Let's consider $N$ interacting Bose particles of mass $m$ confined
in a macroscopic volume $V$. Hamiltonian of the system has a form:
\begin{equation}\label{1.1}
    \hat{H}=\sum_{1\leq j\leq N}\frac{\hat{p}^{2}}{2m}+\sum_{1\leq i<j\leq N}\Phi(|\textbf{r}_{i}-\textbf{r}_{j}|),
\end{equation}
where $\hat{p}$ is the operator of momentum of a particle, the
operator $\Phi(|\textbf{r}_{i}-\textbf{r}_{j}|)$ is an energy of
interaction of two particles. The waves functions must be
symmetrical for any rearrangements of coordinates of any pairs
from $N$ particles and it describes oscillations of Bose liquid.
The space of the collective variables is the suitable multitude of
variables describing collective motion of a system
\cite{vacar7,yukh}. Collective variables are Fourier transform of
fluctuation of density $\triangle
n(\textbf{r})=\sum_{j=1}^{N}\delta(\textbf{r}-\textbf{r}_{j})-N/V$:
\begin{equation}\label{1.2}
    \rho_{\textbf{k}}=\frac{1}{\sqrt{N}}\sum_{j=1}^{N}\exp(-i\textbf{kr})=\rho_{\textbf{k}}^{c}-i\rho_{\textbf{k}}^{s},
\end{equation}
where
\begin{equation}\label{1.3}
    \rho_{\textbf{k}}^{c}=\frac{1}{\sqrt{N}}\sum_{j=1}^{N}\cos(\textbf{kr}), \qquad \rho_{\textbf{k}}^{s}=\frac{1}{\sqrt{N}}\sum_{j=1}^{N}\sin(\textbf{kr}).
\end{equation}
And what's more the correlations have a place:
\begin{equation}\label{1.4}
    \rho_{\textbf{k}}^{\ast}=\rho_{-\textbf{k}} \qquad \Rightarrow \qquad\rho_{\textbf{k}}^{c}= \rho_{\textbf{-k}}^{c}, \qquad \rho_{\textbf{k}}^{s}= -\rho_{\textbf{-k}}^{s}.
\end{equation}
This means, that it is necessary to consider the values
$\rho_{\textbf{k}}$ with indexes $\textbf{k}$ from the half-space
of their possible values only.

The transition from the Cartesian coordinates
($\textbf{r}_{1},...,\textbf{r}_{N}$) to the variables
$\rho_{\textbf{k}}$ is nonunitary because quantity of the
Cartesian coordinates is $3N$ but quantity of
$\rho_{\textbf{k}}$-variables is infinity. Hence superfluous
variables exist among collective variables. The transition to
$\rho_{\textbf{k}}$-representation must be done with the Jacobian
which equalizes a volume of the configuration space $\int
d\textbf{r}_{1}... \int d\textbf{r}_{N}=V^{N}$ to a volume of the
$\rho_{\textbf{k}}$-space:
\begin{equation}\label{1.5}
    V^{N}=\prod'_{\textbf{k}\neq
    0}\int_{-\sqrt{N}}^{\sqrt{N}}d\rho_{\textbf{k}}^{c}\int_{-\sqrt{N}}^{\sqrt{N}}d\rho_{\textbf{k}}^{s}J,
\end{equation}
where the prime at the symbol of multiplication means that
variables $\textbf{k}$ are taken from the half-space only. In a
representation of the new variables (\ref{1.2}) the Hamiltonian
(\ref{1.1}) has the form:
\begin{eqnarray}\label{1.7a}
&&\hat{H}_{\textrm{BZ}}=\sum_{\textbf{k}_{1}\neq
0}\varepsilon(k_{1})\left(\rho_{\textbf{k}_{1}}\frac{\partial}{\partial\rho_{\textbf{k}_{1}}}
-\frac{\partial^{2}}{\partial\rho_{\textbf{k}_{1}}\partial\rho_{-\textbf{k}_{1}}}\right)+
\sum_{\textbf{k}_{1}\neq 0}\sum_{\textbf{k}_{2}\neq
0}^{\textbf{k}_{1}+\textbf{k}_{2}\neq
0}\frac{\varepsilon(\textbf{k}_{1},\textbf{k}_{2})}{\sqrt{N}}\rho_{\textbf{k}_{1}
+\textbf{k}_{2}}\frac{\partial^{2}}{\partial\rho_{\textbf{k}_{1}}\partial\rho_{\textbf{k}_{2}}}
\nonumber\\
&&+\frac{N^{2}}{2V}\nu(0)+\frac{N}{2V}\sum_{\textbf{k}_{1}\neq
0}\nu(k_{1})\left(\rho_{\textbf{k}_{1}}\rho_{-\textbf{k}_{1}}-1\right),
\end{eqnarray}
and its name is Bogoliubov-Zubarev Hamiltonian. However this
operator is non-Hermitian - the first term in the first brackets:
$\rho_{\textbf{k}_{1}}\frac{\partial}{\partial\rho_{\textbf{k}_{1}}}$.
This property of $\hat{H}_{\textrm{BZ}}$ is caused by the the
nonunitarian transition from Cartesian variables to collective
variables $\rho_{\textbf{k}}$.

In the paper \cite{vacar3} the Hermitian form of the Hamiltonian
(\ref{1.7a}) has been obtained. Let the system is described by
the wave function $\psi$ which is normalized in
$\rho_{\textbf{k}}$-representation:
\begin{equation}\label{1.7}
   \prod'_{\textbf{k}\neq
    0}\int_{-\sqrt{N}}^{\sqrt{N}}d\rho_{\textbf{k}}^{c}\int_{-\sqrt{N}}^{\sqrt{N}}d\rho_{\textbf{k}}^{s}J|\psi|^{2}=1.
\end{equation}
Let's introduce the wave functions normalized without the Jacobian
$J$:
\begin{equation}\label{1.8a}
    \bar{\psi}=\psi\sqrt{J}.
\end{equation}
Then Schrodinger equation is written as follows:
\begin{equation}\label{1.8}
    \hat{H}\bar{\psi}=E\bar{\psi}, \qquad
    \hat{H}=J^{1/2}\hat{H}_{\textrm{BZ}}J^{-1/2},
\end{equation}
where the new Hamiltonian $\hat{H}$ must be Hermitian. Proceeding
from this condition we can find the Jacobian $J$. Then we have:
\begin{eqnarray}\label{1.9}
&&\hat{H}=\sum_{\textbf{k}_{1}\neq 0}\varepsilon(k_{1})\left(
-\frac{\partial^{2}}{\partial\rho_{\textbf{k}_{1}}\partial\rho_{-\textbf{k}_{1}}}-\frac{1}{4}\rho_{\textbf{k}_{1}}\frac{\partial
\ln J}{\partial\rho_{\textbf{k}_{1}}}-\frac{1}{2}\right)+
\nonumber \\
\\
&&\sum_{\textbf{k}_{1}\neq 0}\sum_{\textbf{k}_{2}\neq
0}^{\textbf{k}_{1}+\textbf{k}_{2}\neq
0}\frac{\varepsilon(\textbf{k}_{1},\textbf{k}_{2})}{\sqrt{N}}\rho_{\textbf{k}_{1}
+\textbf{k}_{2}}\frac{\partial^{2}}{\partial\rho_{\textbf{k}_{1}}\partial\rho_{\textbf{k}_{2}}}+
\frac{N^{2}}{2V}\nu(0)+\frac{N}{2V}\sum_{\textbf{k}_{1}\neq
0}\nu(k_{1})\left(\rho_{\textbf{k}_{1}}\rho_{-\textbf{k}_{1}}-1\right),\nonumber
\end{eqnarray}
where the Jacobian must be found from the equation:
\begin{equation}\label{1.10}
    \rho_{\textbf{k}_{1}}+\frac{\partial\ln
J}{\partial\rho_{-\textbf{k}_{1}}}-\frac{1}{\sqrt{N}}\sum_{\textbf{k}_{1}\neq
0}^{\textbf{k}_{1}+\textbf{k}_{2}\neq
0}\frac{\textbf{k}_{1}\textbf{k}_{2}}{k^{2}_{1}}\rho_{\textbf{k}_{1}+\textbf{k}_{2}}\frac{\partial
\ln J}{\partial\rho_{\textbf{k}_{2}}}=0.
\end{equation}
Solution of this equation can be written as
\begin{equation}\label{1.10a}
\ln J=\ln C+\sum_{n\geq
2}\frac{(-1)^{n-1}}{n(n-1)(\sqrt{N})^{n-2}}\sum_{\textbf{q}_{1}\neq
0}\ldots\sum_{\textbf{q}_{n}\neq
0}\rho_{\textbf{q}_{1}}...\rho_{\textbf{q}_{n}}.
\end{equation}
The constant $C$ can be found from the condition (\ref{1.5}). I
our comprehension the equation (\ref{1.8}) with the Hamiltonian
(\ref{1.9}) is \emph{motion equation} in
$\rho_{\textbf{k}}$-space. The equation (\ref{1.10}) is
\emph{constraint equation} in this space. In Cartesian coordinate
we have a discrete system from $N$ particles but in collective
coordinates the system is regarded as continuum. The constraints
equation removes the superfluous degrees of freedom.

Normal motion and superfluid motion can be separated in the
Schrodinger equation \cite{vacar1}. The wave function of a ground
state describes zero-point oscillations of Bose liquid and has an
exponential form $e^{U}$. The whole wave function (with excited
states) has a form:
\begin{equation}\label{1.11}
    \bar{\psi}=e^{U}\varphi.
\end{equation}
Then let's rewrite an energy of a liquid as
\begin{equation}\label{1.12}
E=E_{0}+E-E_{0}\equiv E_{0}+E_{ext},
\end{equation}
where $E_{0}$ - is a ground state energy (energy of
superfluid motion with wave function $e^{U}$), $E_{ext}$ - is the
energy of excitation (energy of normal motion with wave function
$\varphi$). Then the equation (\ref{1.8}) divides into the set of equations
describing superfluid motion
\begin{eqnarray}\label{1.13}
&&-\sum_{\textbf{k}_{1}\neq 0}\varepsilon(k_{1})\left[
\frac{\partial^{2}U}{\partial\rho_{\textbf{k}_{1}}\partial\rho_{-\textbf{k}_{1}}}+\frac{\partial
U}{\partial\rho_{\textbf{k}_{1}}}\frac{\partial
U}{\partial\rho_{-\textbf{k}_{1}}}\right]\nonumber\\
&&+\sum_{\textbf{k}_{1}\neq 0}\sum_{\textbf{k}_{2}\neq
0}\frac{\varepsilon(\textbf{k}_{1},\textbf{k}_{2})}{\sqrt{N}}\rho_{\textbf{k}_{1}
+\textbf{k}_{2}}\left[\frac{\partial
U}{\partial\rho_{\textbf{k}_{1}}\partial\rho_{\textbf{k}_{2}}}+\frac{\partial
U}{\partial\rho_{\textbf{k}_{1}}}\frac{\partial
U}{\partial\rho_{\textbf{k}_{2}}}\right]
\\
&&+\sum_{\textbf{k}_{1}\neq
0}\left[\frac{N}{2V}\nu(k_{1})\rho_{\textbf{k}_{1}}\rho_{-\textbf{k}_{1}}-\frac{1}{4}\rho_{\textbf{k}_{1}}\frac{\partial
\ln J}{\partial\rho_{\textbf{k}_{1}}}\varepsilon(k_{1})\right]
=E_{0}+\sum_{\textbf{k}\neq
0}\left[\frac{1}{2}\varepsilon(k)+\frac{N}{2V}\nu(k)\right]-\frac{N^{2}}{2V}\nu(0),\nonumber
\end{eqnarray}
and normal motion
\begin{eqnarray}\label{1.14}
&&-\sum_{\textbf{k}_{1}\neq 0}\varepsilon(k_{1})\left[
\frac{\partial^{2}\varphi}{\partial\rho_{\textbf{k}_{1}}\partial\rho_{-\textbf{k}_{1}}}+2\frac{\partial
U}{\partial\rho_{\textbf{k}_{1}}}\frac{\partial
\varphi}{\partial\rho_{-\textbf{k}_{1}}}\right]+
\nonumber \\
&&\sum_{\textbf{k}_{1}\neq 0}\sum_{\textbf{k}_{2}\neq
0}\frac{\varepsilon(\textbf{k}_{1},\textbf{k}_{2})}{\sqrt{N}}\rho_{\textbf{k}_{1}+
\textbf{k}_{2}}\left[\frac{\partial
\varphi}{\partial\rho_{\textbf{k}_{1}}\partial\rho_{\textbf{k}_{2}}}+2\frac{\partial
U}{\partial\rho_{\textbf{k}_{1}}}\frac{\partial
\varphi}{\partial\rho_{\textbf{k}_{2}}}\right]=E_{ext}(k)\varphi
\end{eqnarray}

The wave function $\psi$ must be eigenfunction of a momentum
operator \cite{vacar1,vacar2}:
\begin{equation}\label{1.15}
    \hat{\textbf{P}}=-\sum_{\textbf{k}_{1}\neq
    0}\hbar\textbf{k}_{1}\rho_{\textbf{k}_{1}}\frac{\partial}{\partial\rho_{\textbf{k}_{1}}},
\qquad \hat{\textbf{P}}\psi=\textbf{P}\psi.
\end{equation}
The wave function $\bar{\psi}$ is eigenfunction of the momentum
operator too:
\begin{equation}\label{1.16}
    J^{1/2}\hat{\textbf{P}}J^{-1/2}\bar{\psi}=\textbf{P}\bar{\psi}\Longrightarrow\hat{\textbf{P}}\bar{\psi}=\textbf{P}\bar{\psi}.
\end{equation}
The proof of this fact is in Appendix \ref{A}. Then solution of
the equations (\ref{1.13}) and (\ref{1.14}) must obey the
condition:
\begin{equation}\label{1.17}
    \hat{\textbf{P}}e^{U}=0\cdot e^{U}, \qquad \hat{\textbf{P}}\varphi=\hbar(\textbf{k}_{1}+\ldots+\textbf{k}_{n})\varphi.
\end{equation}
This means that a center of the system's mass rests and the
function $\varphi$ describes excited states of a system, which are
characterized with conserving energy $E_{ext}(k)$ and with
conserving momentum $\hbar(\textbf{k}_{1}+\ldots+\textbf{k}_{n})$.
Thus the function $\varphi$ describes state with $n$ collective
excitations. If an interaction of particles $\nu(k)=0$ then the
wave function $\psi$ is a wave function of free bosons
$\psi_{\nu=0}=1/\sqrt{V^{N}}$. A proof of this statement is in
Appendix \ref{B}.

In the total case a solution of the equation (\ref{1.13}) has a
form of correlation series expasion\cite{vacar1,vacar2}:
\begin{equation}\label{1.20}
    U=\sum_{\textbf{k}\neq 0}f(k)\rho_{\textbf{k}}\rho_{-\textbf{k}}+
    \sum_{\textbf{k}\neq 0}\sum_{\textbf{q}\neq 0}^{\textbf{k}+\textbf{q}\neq 0}
    \frac{g(\textbf{k},\textbf{q})}{\sqrt{N}}\rho_{\textbf{q}-\textbf{k}}\rho_{\textbf{k}}\rho_{-\textbf{q}}+\ldots
\end{equation}
Analogous expansion for the function $\varphi$ of state with one
CE (as $\hat{\textbf{P}}\varphi=\hbar\textbf{k}\varphi$) has a
form:
\begin{equation}\label{1.21}
    \varphi=\rho_{-\textbf{k}}+
    \sum_{\textbf{q}\neq 0}^{\textbf{k}+\textbf{q}\neq 0}
    \frac{L(\textbf{k},\textbf{q})}{\sqrt{N}}\rho_{\textbf{q}-\textbf{k}}\rho_{-\textbf{q}}+\ldots
\end{equation}

Unfortunately no small parameter exist in these expansions. And
what's more, calculation of higher terms of the expansions is very
difficult mathematically. In order to solve this problem we have
to use some model considerations and approximations.

We have to make a little remark hear. Under \emph{collective
excitations} we understand quants of collective motions of
macroscopic group of particles, that is with motion of system as a
whole. For example: phonon, plasmon, magnon. Under
\emph{quasi-particle} we understand a particle interacting with
its environment or external field (''dressed'' particle). The
quasi-particles are characterized by effective mass and they are
interacting with effective (screened) potential. The examples of
quasi-particle are conduction electron, polaron, Cooper pair.

\section{Random Phase Approximation}\label{RPA}
\subsection{The random phases as zeroth approximation}\label{subRPA1}

In this section we will shortly formulate zeroth approximation to
our problem - RPA or harmonic approximation. The Jacobian and the
wave function of a ground state we write in Gauss form:
\begin{equation}\label{2.1}
\ln J=\ln C-\frac{1}{2}\sum_{\textbf{k}\neq
0}\rho_{\textbf{k}}\rho_{-\textbf{k}}.
\end{equation}
\begin{equation}\label{2.2}
    U=\sum_{\textbf{k}\neq
    0}f(k)\rho_{\textbf{k}}\rho_{-\textbf{k}}.
\end{equation}
Substituting these expressions in the equation (\ref{1.13}) and
neglecting by powers of $\rho_{\textbf{k}}$ higher than second
power we have:
\begin{eqnarray}\label{2.3}
&&\sum_{\textbf{k}\neq
0}\left[-4\varepsilon(k)f^{2}(k)+\frac{1}{4}\varepsilon(k)+\frac{N}{2V}\nu(k)\right]\rho_{\textbf{k}}\rho_{-\textbf{k}}=
\nonumber\\
&&E_{0}-\frac{N^{2}}{2V}\nu(0)+\sum_{\textbf{k}\neq
0}\left[2\varepsilon(k)f(k)+\frac{1}{2}\varepsilon(k)+\frac{N}{2V}\nu(k)\right].
\end{eqnarray}
From this equation we can obtain unknown function $f$:
\begin{equation}\label{2.4}
    f(k)=-\frac{1}{4}\sqrt{1+\frac{2N}{V}\frac{\nu(k)}{\varepsilon(k)}},
\end{equation}
and corresponding energy:
\begin{eqnarray}\label{2.5}
E_{0}=\sum_{\textbf{k}\neq
0}\frac{1}{2}\sqrt{\varepsilon(k)^{2}+\frac{2N}{V}\nu(k)\varepsilon(k)}
-\sum_{\textbf{k}\neq
0}\left[\frac{1}{2}\varepsilon(k)+\frac{N}{2V}\nu(k)\right]+\frac{N^{2}}{2V}\nu(0).
\end{eqnarray}
It is necessary to notice that the solutions (\ref{2.4},
\ref{2.5}) correspond to the shortened Hamiltonian:
\begin{eqnarray}\label{2.5a}
&&\hat{H}_{\textrm{RPA}}=\sum_{\textbf{k}_{1}\neq
0}\varepsilon(k_{1})\left(
-\frac{\partial^{2}}{\partial\rho_{\textbf{k}_{1}}\partial\rho_{-\textbf{k}_{1}}}+\frac{1}{4}\rho_{\textbf{k}_{1}}\rho_{-\textbf{k}_{1}}-\frac{1}{2}\right)
\nonumber\\
&&+\frac{N^{2}}{2V}\nu(0)+\frac{N}{2V}\sum_{\textbf{k}_{1}\neq
0}\nu(k_{1})\left(\rho_{\textbf{k}_{1}}\rho_{-\textbf{k}_{1}}-1\right),
\end{eqnarray}
where, unlike the complete Hamiltonian (\ref{1.9}), the harmonic
terms is kept only.

A structure factor is the most important characteristic of liquids
\cite{vacar7}:
\begin{equation}\label{2.6}
    S(k)=\langle\rho_{\textbf{k}}\rho_{-\textbf{k}}\rangle,
\end{equation}
where the mean value $\langle\rangle$ is calculated with a ground
state. This function can be obtained by using the virial theorem:
\begin{equation}\label{2.7}
\frac{\delta E}{\delta\nu(k)}=\left\langle\frac{\delta
\hat{H}}{\delta\nu(k)}\right\rangle\Rightarrow
S(k)=\frac{1}{\sqrt{1+\frac{2N}{V}\frac{\nu(k)}{\varepsilon(k)}}}=-\frac{1}{4f(k)}.
\end{equation}

For atoms of helium the potential of interaction is very like to
Lennard-Jones potential in the region $r>2.5A$. In the remaining
region the interaction is known badly and various adaption
functions have be used here. In order to get over this difficulty
the approach with structure factor has been developed in some
papers where the potential of interaction is unknown function and
the structure factor is taken from experiment data. This means,
that \emph{the perturbation theory must be constructed so we must
obtain the same structure factor in each approximation}. On the
present step of calculation the potential of interaction is
obtained from (\ref{2.7}) in a form:
\begin{equation}\label{2.9}
\nu(k)=\frac{V}{2N}\varepsilon(k)\left[\frac{1}{S^{2}(k)}-1\right].
\end{equation}

In order to obtain dispersion curve of a CE we assume that
$\varphi=\rho_{-\textbf{k}}$. Substituting its and the function
(\ref{2.2}) in the equation (\ref{1.14}) we have:
\begin{equation}\label{2.12}
E_{ext}^{RPA}=\frac{\varepsilon(k)}{S(k)}=\sqrt{\varepsilon(k)^{2}+\frac{2N}{V}\nu(k)\varepsilon(k)}.
\end{equation}
This expression is known as Feynman formula or Bogoliubov specter.
The formula (\ref{2.12}) is a starting-point in order to obtain
the dispersion curve of a CE which is consistent with the experimental
specter.

\subsection{Formulation of the oscillator model}\label{subRPA2}

The Hamiltonian (\ref{2.5a}) is sum of separate terms where each
of them is characterized by own wave vector $\textbf{k}$. This
fact means that infinity number of independent motions can be in a
system, and every of them is characterized by wave vector
$\textbf{k}$ and some energy $\xi(k)$. For the each independent
motion Schrodinger equation is written as follows:
\begin{eqnarray}\label{2.13}
-\frac{\hbar^{2}}{2(m/k^2)}\frac{\partial^{2}\bar{\psi}}{\partial\rho_{\textbf{k}}\partial\rho_{-\textbf{k}}}
+\left[\frac{1}{4}\varepsilon(k)+\frac{N}{2V}\nu(k)\right]\rho_{\textbf{k}}\rho_{-\textbf{k}}\bar{\psi}=\xi(k)\bar{\psi}.
\end{eqnarray}
This equation is similar to an equation of a harmonic oscillator, where
\begin{eqnarray}\label{2.14}
\frac{\hbar^{2}k^2}{2m}\equiv\frac{\hbar^{2}}{2M}\Rightarrow
M\equiv\frac{m}{k^{2}}, \qquad
\frac{1}{4}\varepsilon(k)+\frac{N}{2V}\nu(k)\equiv\frac{1}{2}M\omega^{2}.
\end{eqnarray}
$\xi(k)$ is energy of a oscillator with the wave vector
$\textbf{k}$:
\begin{equation}\label{2.15}
\xi(k)=\hbar\omega(k)\left(\frac{1}{2}+n\right)=\frac{\varepsilon(k)}{S(k)}\left(\frac{1}{2}+n\right),
\end{equation}
and the ground state energy of a system is
\begin{eqnarray}\label{2.16}
E_{0}=\sum_{\textbf{k}\neq 0}\frac{1}{2}\hbar\omega(k)
-\sum_{\textbf{k}\neq
0}\left[\frac{1}{2}\varepsilon(k)+\frac{N}{2V}\nu(k)\right]+\frac{N^{2}}{2V}\nu(0).
\end{eqnarray}
The wave function of a ground state has the form:
\begin{equation}\label{2.17}
\bar{\psi}\sim\exp\left(-\frac{\rho_{\textbf{k}}\rho_{-\textbf{k}}}{2l^{2}}\right)=e^{f(k)\rho_{\textbf{k}}\rho_{-\textbf{k}}},
\qquad l^{2}=\frac{\hbar}{M\omega}
\end{equation}
where $l$ is a oscillator length. The structure factor can be
found from the virial theorem for an oscillator:
\begin{equation}\label{2.18}
\frac{M\omega^{2}}{2}\langle\rho_{\textbf{k}}\rho_{-\textbf{k}}\rangle=\frac{\xi(k)}{2}\Rightarrow
S=\frac{\varepsilon(k)}{\hbar\omega(k)}[1+2\langle n\rangle],
\end{equation}
and the state $n=0$ corresponds to a ground state. A structure
factor $S(k)$ is dispersion of a oscillator with a wave vector
$\textbf{k}$.

On the assumption of the aforesaid we can formulate the oscillator
model of Bose liquid. A quantum liquid represents a totality of
harmonic oscillators. Each oscillator is a vibration mode of
density of liquid. A ground state energy of a system is a sum
of energies of ground states of these oscillators. Availability in
a system $n$ collective excitations corresponds to a state with
$n$ excited oscillators. The oscillator model is a model of
continuum medium and it is incorrect on distances between
neighboring atoms. The calculation of short-range correlations is
very difficult because all anharmonic terms in the Hamiltonian
(\ref{1.9}) and in the Jacobian (\ref{1.10a}) must be calculated.
Moreover if we shall try to calculate them with help ordinary
perturbation theory then we shall have divergent integrals. This
difficulty appears due contribution of above-mentioned superfluous
degrees of freedom. In the next sections we shall formulate the
method of calculation of the anharmonic terms remaining in limits
of the oscillator model.

\section{The spectrum of collective excitation}\label{Spectrum}
\subsection{Decay of collective excitation}\label{subSpectrum1}

\begin{figure}[]
\centerline{\includegraphics[width=8.0cm]{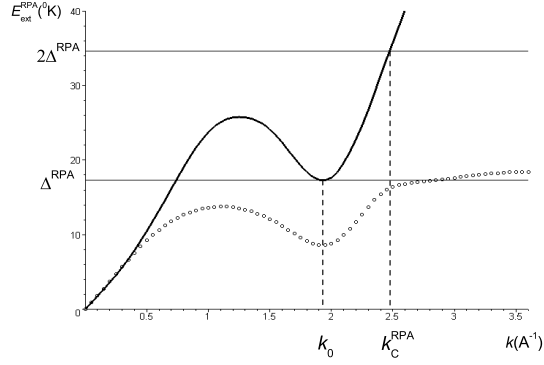}} \caption{The
spectrum of CE calculated with the Feynman formula (solid line)
and measured in experiment (dotted line). Decay of excitation on
two rotons is possible in the region $k>k_{\textrm{C}}^{RPA}$
where $E_{ext}^{RPA}(k)>2\Delta^{RPA}$. The experimental curve has a
"shelf" at $k>2.5A^{-1}$ where decay of CE into two rotons is
observed.} \label{fig1}
\end{figure}

The wave function of a excited state $\varphi=\rho_{-\textbf{k}}$
leads to the well known Feynman formula (\ref{2.12}). We can see
on Fig.\ref{fig1} that the dispersion curve in RPA coincides with
the experimental dispersion curve in the limit $k\rightarrow 0$
only. Its basic contrast to the real spectrum is that that Feynman
formula doesn't take into account decay of CE and the end of a
dispersion curve accordingly.

\begin{figure}[tbp]
\centerline{\includegraphics[width=15.0cm]{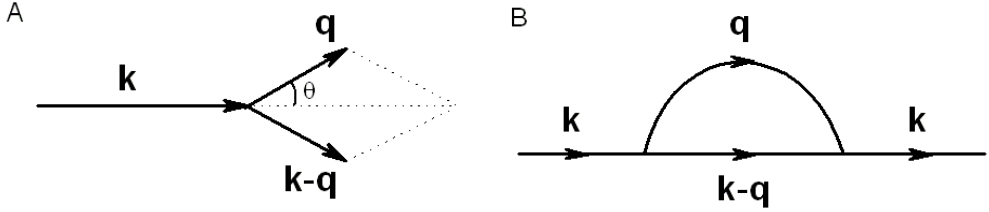}} \caption{The
processes described by the second term in the wave function
(\ref{1.21}). The process (A) is a real decay, the process (B) is
a virtual decay of CE with momentum $\hbar\textbf{k}$ into two
excitations with momentums $\hbar(\textbf{k}-\textbf{q})$ and
$\hbar\textbf{q}$.} \label{fig2}
\end{figure}

Let's consider the wave function of CE (\ref{1.21}) with momentum
$\hbar\textbf{k}$ where $L(\textbf{k},\textbf{q})$ is the unknown
function. In our interpretation the second term in (\ref{1.21})
describes a process of decay of CE with momentum $\hbar\textbf{k}$
and energy $E_{ext}(k)$ into two excitations with momentums
$\hbar(\textbf{k}-\textbf{q})$ and $\hbar\textbf{q}$ and with
energies $E_{ext}(|\textbf{k}-\textbf{q}|)$ and $E_{ext}(q)$
accordingly. Decay of CE can occur if energy is conserved:
\begin{equation}\label{3.2}
    E_{ext}(k)=E_{ext}(|\textbf{k}-\textbf{q}|)+E_{ext}(q).
\end{equation}
Such situation corresponds to the case (A) on Fig.\ref{fig2}. This
process causes damping of a CE and the end of the dispersion curve
\cite{lif}. In those parts of the energy spectrum where the
equality (\ref{3.2}) isn't true, CE is stable. In this case the
term in (\ref{1.21}) with the multiplier
$\rho_{\textbf{q}-\textbf{k}}\rho_{-\textbf{q}}$ describes
processes of virtual decay and creation of CE. This causes trivial
renormalization of the energy spectrum of CE. This situation
corresponds to the case (B) on Fig.\ref{fig2}. Higher
approximations in (\ref{1.21}) describe decays of CE on three and
more excitations. We suppose that such processes is improbable and
we shall not consider them in this paper.

Substituting the wave function (\ref{1.21}) in the equation
(\ref{1.14}) we can obtain the function
\begin{equation}\label{3.3}
L(\textbf{k},\textbf{q})=\frac{\varepsilon(\textbf{k},\textbf{q})S^{-1}(q)}{E_{ext}(k)-E_{ext}^{RPA}(|\textbf{k}-\textbf{q}|)-E_{ext}^{RPA}(q)},
\end{equation}
where $E_{ext}^{RPA}(q)$ is spectrum of CE in RPA - the formula
(\ref{2.12}). And besides we can obtain the dispersion law of CE
in a form:
\begin{eqnarray}\label{3.4}
E_{ext}(k)&=&E_{ext}^{RPA}(k)+\frac{2}{N}\sum_{\textbf{q}\neq 0}L\varepsilon(\textbf{k}-\textbf{q},\textbf{q})\nonumber\\
&=&E_{ext}^{RPA}(k)-\frac{2}{N}\sum_{\textbf{q}\neq
0}\frac{\varepsilon(\textbf{k},\textbf{q})\varepsilon(\textbf{k}-\textbf{q},\textbf{q})S^{-1}(q)}{E_{ext}^{RPA}(|\textbf{k}-\textbf{q}|)+E_{ext}^{RPA}(q)-E_{ext}(k)}.
\end{eqnarray}
This equation is an integral equation for unknown function
$E_{ext}(k)$. The first term corresponds to the term
$\rho_{-\textbf{k}}$ in the wave function (\ref{1.21}) and it is
Bogoliubov-Feynman spectrum. The second term has a singularity in
those points $q$ where the equality is executed:
\begin{equation}\label{3.5}
    E_{ext}(k)=E_{ext}^{RPA}(|\textbf{k}-\textbf{q}|)+E_{ext}^{RPA}(q).
\end{equation}
This equation likes the expression (\ref{3.2}) and it is condition
of decay of CE. Inexactitude of this condition is in that that a
real CE $E_{ext}(k)$ decays on excitations with a spectrum
calculated in RPA: $E_{ext}^{RPA}(q)$. However in the region of
large $q$ the theoretical dispersion law of CE differs from the
real law $E_{ext}(q)$ essentially. We shall get over this
difficulty later. For the present we shall work with the condition
(\ref{3.5}).

CE is stable at those $k$ where the equation (\ref{3.5}) hasn't
solutions. Damping of CE appears in a threshold of decay
$k=k_{\textrm{C}}$. At the point $k=k_{\textrm{C}}$ a solution of
the equation (\ref{3.5}) appears at first, and the right sides of
the equations (\ref{3.2}) and (\ref{3.5}) have extremum \cite{lif}
as a function of $q$. We can see on Fig.\ref{fig1} that the
extremum for the function $E_{ext}^{RPA}(q)$ is the roton minimum.
In its neighborhood the spectrum of collective excitations has a
view:
\begin{eqnarray}\label{3.6}
&&E_{ext}^{RPA}(q)=\Delta^{RPA}+\frac{\hbar^{2}}{2\mu}(q-k_{0})^{2},\\
&&E_{ext}^{RPA}(|\textbf{k}-\textbf{q}|)=\Delta^{RPA}+\frac{\hbar^{2}}{2\mu}(|\textbf{k}-\textbf{q}|-k_{0})^{2},\label{3.6a}
\end{eqnarray}
where $\Delta^{RPA}=17.3^{0}K$ is the roton gap in RPA (in degrees
of Kelvin), $\mu=0.09m$ is effective mass of a roton in RPA,
$k_{0}$ is momentum corresponding to the roton gap (in angstroms).
We can suppose that $k_{0}=k_{0}^{RPA}=1.93 A^{-1}$ in all
approximations. The condition (\ref{3.5}) is executed when
$E_{ext}(k)=2\Delta^{RPA}$. Then the underintegral function in the spectrum
(\ref{3.4}) at $k=k_{C}$ has a singularity in the point $q=q_{0}$.
This means decay of CE with momentum $\hbar k_{\textrm{C}}$ into
two rotons with momentums $\hbar k_{0}$ each. Hence we have the
connection $k_{\textrm{C}}=2k_{0}\cos\theta_{c}$ where
$2\theta_{c}$ is a angle of recession of two rotons.

Let's evaluate the integral in (\ref{3.4}) in the
point $k=k_{\textrm{C}}$. Let's consider asymptotics of the
underintegral function at $q\rightarrow\infty$ where
$S(q)\rightarrow 1$. If a vector $\textbf{k}$ is directed along the axe
$Oz$ then $\textbf{kq}=kq\cos\theta$. Let's use the expressions
(\ref{3.6}) and (\ref{3.6a}) in the denominator of (\ref{3.4})
as they determine the singularity at $q=k_{0}$, and we have $|\textbf{k}-\textbf{q}|\approx
q$. Then the integral in (\ref{3.4}) is written in a form:
\begin{eqnarray}
&&\frac{2}{N}\frac{V}{(2\pi)^{3}}\left(\frac{\hbar^{2}}{2m}\right)^{2}
%\left
(\int\frac{(qk\cos\theta)^{2} q^{2}dq\sin\theta d\theta
d\varphi}{2\Delta^{RPA}-E_{ext}+\frac{\hbar^{2}}{\mu}(q-k_{0})^{2}}
\nonumber\\
&&-\int\frac{qk\cos\theta q^{2}q^{2}dq\sin\theta d\theta
d\varphi}{2\Delta^{RPA}-E_{ext}+\frac{\hbar^{2}}{\mu}(q-k_{0})^{2}}\label{3.7}
%\right
)\\
&&=\frac{2}{N}\frac{V}{(2\pi)^{2}}\left(\frac{\hbar^{2}}{2m}\right)^{2}\frac{2}{3}k^{2}
\int\frac{q^{4}dq}{2\Delta^{RPA}-E_{ext}+\frac{\hbar^{2}}{\mu}(q-k_{0})^{2}}.\label{3.7a}
\end{eqnarray}
In the second integral in (\ref{3.7}) integration by $\theta$
gives zero. Taking into consideration
$E_{ext}(k_{\textrm{C}})=2\Delta^{RPA}$ we can rewrite
(\ref{3.7a}) as
\begin{equation}\label{3.7b}
\frac{2}{N}\frac{V}{(2\pi)^{2}}\left(\frac{\hbar^{2}}{2m}\right)^{2}\frac{2}{3}k^{2}\frac{\mu}{\hbar^{2}}
\int_{0}^{\infty}\frac{q^{4}}{(q-k_{0})^{2}}dq\rightarrow\infty
\end{equation}
The underintegral function has a pole in the point
$k_{0}$ that results in some complex addendum in an
energy of CE: $E_{ext}-i/\tau$. This means damping and limited
life time $\sim\tau$ of CE. Moreover we can see that
\emph{ultraviolet divergence} takes place. Hence the second term
in the formula (\ref{3.4}) for the spectrum of CE aspires
to infinity: $E_{ext}(k)=E_{ext}^{RPA}(k)+\infty$. The nature of
this divergence lies in the following.

The constraint equation (\ref{1.10}) removes the superfluous
degrees of freedom appeared at nonunitarian transition to collective
coordinates. These degrees of freedom are connected with continuum
representation of primarily discreet medium. Each superfluous degree of freedom gives
contribution to energy. Neglect of all anharmonic terms in the
Jacobian (\ref{1.10a}) and in the Hamiltonian (\ref{1.9})
doesn't give possibility to delete this nonphysical
energy. Contribution of the superfluous
degrees of freedom results in infinity in the integral
(\ref{3.4}).

The method of removal of the divergence is cut-off of the integral
(\ref{3.4}) to some unknown value of a wave vector $q=q_{m}$. By
analogy with \cite{lif,pit} let's introduce new variables $q'_{z}$
ш $q'_{\rho}$ in accordance with the definition:
\begin{eqnarray}\label{3.8}
&&q_{x}=(k_{0}\sin\theta_{c}+q'_{\rho})\cos\varphi \qquad q_{y}=(k_{0}\sin\theta_{c}+q'_{\rho})\sin\varphi\nonumber\\
&&q_{z}=k_{0}\cos\theta_{c}+q'_{z} \qquad  d^{3}q=q^{2}dq\sin\theta d\theta
d\varphi=(k_{0}\sin\theta_{c}+q'_{\rho})dq'_{\rho}dq'_{z}d\varphi.
\end{eqnarray}
In a neighborhood of the
threshold of decay $k=k_{\textrm{C}}$ (where
$E_{ext}\rightarrow 2\Delta^{RPA}$) the underintegral function in
(\ref{3.4}) has a pole in the point [$q=k_{0}$,
$\theta=\theta_{c}$]. We have
$|q'_{z}|\ll k_{0}$ and $|q'_{\rho}|\ll k_{0}$ in a neighborhood of
the singularity $q\rightarrow k_{0}, \theta\rightarrow\theta_{c}$.
Then we can write expansions near the singularity:

\begin{eqnarray}\label{3.9}
&&d^{3}q\approx k_{0}\sin\theta_{c} dq'_{\rho}dq'_{z}d\varphi ,\qquad q\approx k_{0}+q'_{z}\cos\theta_{c}+q'_{\rho}\sin\theta_{c}\nonumber\\
&&|\textbf{k}-\textbf{q}|\approx
k_{0}+q'_{\rho}\sin\theta_{c}-q'_{z}\cos\theta_{c}, \qquad \textbf{k}\textbf{q}\approx
kk_{0}\cos\theta_{c}+O(q'_{\rho},q'_{z})\nonumber\\
&&(\textbf{k}-\textbf{q})\textbf{q}\approx
k_{0}^{2}\cos2\theta_{c}+O(q'_{\rho},q'_{z}).
\end{eqnarray}
In the limit $q\rightarrow k_{0}$ energies of products of decay
have a view:
\begin{eqnarray}\label{3.9a}
&&E_{ext}^{RPA}(q)=\Delta^{RPA}+\frac{\hbar^{2}}{2\mu}(q'_{z}\cos\theta_{c}+q'_{\rho}\sin\theta_{c})^{2},\nonumber\\
&&E_{ext}^{RPA}(|\textbf{k}-\textbf{q}|)=\Delta^{RPA}+\frac{\hbar^{2}}{2\mu}(q'_{\rho}\sin\theta_{c}-q'_{z}\cos\theta_{c})^{2}.
\end{eqnarray}

Let's consider the spectrum of CE (\ref{3.4}) in some
neighborhood of the threshold of decay $k\rightarrow
k_{\textrm{C}}$. Decomposing the underintegral function near the
pole $q\rightarrow k_{0}$, $\theta\rightarrow\theta_{c}$ using the
expansions (\ref{3.9}) and (\ref{3.9a}) we have::
\begin{eqnarray}
&&-\frac{2}{N}\sum_{\textbf{q}\neq
0}\frac{\varepsilon(\textbf{k},\textbf{q})
\varepsilon(\textbf{k}-\textbf{q},\textbf{q})S^{-1}(q)}{E_{ext}^{RPA}(|\textbf{k}-\textbf{q}|)+E_{ext}^{RPA}(q)-E_{ext}(k)}\label{3.10a}\\
&&\rightarrow
-\frac{2}{N}\frac{V}{(2\pi)^{2}}\left(\frac{\hbar^{2}k^{2}_{0}}{2m}\right)^{2}\frac{\mu}{\hbar^{2}}S^{-1}(k_{0})\cos2\theta_{c}\cos\theta_{c}\sin\theta_{c}
\nonumber\\
&&\times
k\int\frac{dq'_{z}dq'_{\rho}}{\frac{\mu}{\hbar^{2}}(2\Delta^{RPA}-E_{ext})
+\left[(q'_{\rho})^{2}\sin^{2}\theta_{c}+(q'_{z})^{2}\cos^{2}\theta_{c}\right]}\label{3.10b}.
\end{eqnarray}
However the
integrals (\ref{3.10a}) and (\ref{3.10b}) are not equal because
transition to the limit $q\rightarrow k_{0},
\qquad\theta\rightarrow\theta_{c}$ has been done in the underintegral
function. Integration by all $q$-space results to infinity. Then we must introduce some cut-off
parameter $\rho$ so as the integral (\ref{3.10b}) is finite
on the one hand, and the integrals
(\ref{3.10a}) and (\ref{3.10b}) are approximately equal in a neighborhood of the point
$q=k_{0}$: $k_{0}-\rho< q< k_{0}+\rho$ on the other hand.

Now let's introduce the polar coordinates:
\begin{eqnarray}\label{3.11}
&&q'_{\rho}\sin\theta_{c}=\rho\cos\psi\nonumber\\
&&q'_{z}\cos\theta_{c}=\rho\sin\psi\nonumber\\
&&dq'_{\rho}dq'_{z}=\frac{\rho d\rho
d\psi}{\cos\theta_{c}\sin\theta_{c}}.
\end{eqnarray}
Then the integral (\ref{3.10b}) is reduced to the form:
\begin{eqnarray}\label{3.12}
&&-\frac{2}{N}\frac{V}{2\pi}\varepsilon^{2}(k_{0})\frac{\mu\cos2\theta_{c}}{\hbar^{2}}S^{-1}(k_{0})
k\int_{0}^{\rho}\frac{\rho
d\rho}{\frac{\hbar^{2}}{\mu}(2\Delta^{RPA}-E_{ext})+\rho^{2}}\\
&&=\frac{V}{N}\frac{\varepsilon^{2}(k_{0})\mu\cos2\theta_{c}}{2\pi\hbar^{2}S(k_{0})}
k\ln\left[\frac{(2\Delta^{RPA}-E_{ext})\mu/\hbar^{2}}{(2\Delta^{RPA}-E_{ext})\mu/\hbar^{2}+\rho^{2}}\right],\nonumber
\end{eqnarray}
We can see that the final result depends on the unknown parameter
$\rho$ which determines a integration domain near
the point $q=k_{0}$. Then $\rho$ must satisfy the requirements:

\begin{enumerate}
    \item The point $k_{\textrm{C}}$ is the threshold of decay:
$E_{ext}(k_{\textrm{C}})=2\Delta^{RPA}$ \label{i}
    \item On Fig.\ref{fig1} we can see that Feynman formula
$E_{ext}^{RPA}(k)=\varepsilon(k)/S(k)$ in the limit $k\rightarrow
0$ gives a sound mode $c\hbar k$ with a correct sound velocity $c$.
Hence \emph{the total spectrum (\ref{3.4}) must have the same asymptotic}:
\begin{equation}\label{3.13}
    E_{ext}(k\rightarrow 0)=E_{ext}^{RPA}(k\rightarrow 0)=c\hbar k.
\end{equation}
The second term of the expression (\ref{3.4}) in the limit
$k\rightarrow k_{\textrm{C}}$ is the integral (\ref{3.10b}). We
must join these two asymptotics in a single expression which is a sought dispersion curve.
Hence the parameter $\rho$ is
function of variable $k$: $\rho=\rho(k)$.\label{ii}
    \item The parameter $\rho$ must ensure
approximate equality of the integrals (\ref{3.10a}) and
(\ref{3.10b}) in a neighborhood of the point $q=k_{0}$. We have
decomposed the underintegral function in (\ref{3.10a}) near the point
$q=k_{0}$ in power series of $\rho/k_{0}$:
\begin{eqnarray}\label{3.13a}
&&q\approx
k_{0}+q'_{z}\cos\theta_{c}+q'_{\rho}\sin\theta_{c}\approx
k_{0}\left(1+\rho/k_{0}\right),\nonumber\\
&&|\textbf{k}-\textbf{q}|\approx
k_{0}+q'_{\rho}\sin\theta_{c}-q'_{z}\cos\theta_{c}\approx
k_{0}\left(1+\rho/k_{0}\right),\nonumber\\
&&k_{0}\sin\theta_{c}+q'_{\rho}\approx
k_{0}\sin\theta_{c}\left(1+\frac{\rho}{k_{0}\sin\theta_{c}}\right)
\Rightarrow\frac{\rho}{k_{0}\sin\theta_{c}}\ll 1.
\end{eqnarray}
Hence $\rho(k)$ must satisfy the inequality (\ref{3.13a}).\label{iii}
    \item The spectrum of CE (\ref{3.4}) must be determined with
the parameter $\rho(k)$ by self-consistent way.\label{iiii}
\end{enumerate}

In order to meet the requirements (\ref{i}-\ref{iiii}) we must
assume that
\begin{equation}\label{3.14}
\frac{(2\Delta^{RPA}-E_{ext})\mu/\hbar^{2}}{(2\Delta^{RPA}-E_{ext})\mu/\hbar^{2}+\rho^{2}}=
\frac{2\chi\Delta^{RPA}-E_{ext}}{2\chi\Delta^{RPA}},
\end{equation}
then
\begin{equation}\label{3.14a}
\rho^{2}(k)=\frac{\mu}{\hbar^{2}}\frac{E_{ext}(k)\left[2\Delta^{RPA}-E_{ext}(k)\right]}{2\chi\Delta^{RPA}-E_{ext}(k)},
\end{equation}
where the parameter $\chi$ is determined by the condition
(\ref{i}) $E_{ext}(k_{\textrm{C}})=2\Delta^{RPA}$:
\begin{equation}\label{3.14b}
\chi=\left[1-\exp\left(\frac{2\Delta^{RPA}-E^{RPA}_{ext}(k_{\textrm{C}})}{\alpha
k_{\textrm{C}}}\right)\right]^{-1}.
\end{equation}
As a result the dispersion equation for CE
(\ref{3.4}) has a form:
\begin{equation}\label{3.15}
E_{ext}(k)=E_{ext}^{RPA}(k)+\alpha
k\ln\left[\frac{2\chi\Delta^{RPA}-E_{ext}(k)}{2\chi\Delta^{RPA}}\right],
\end{equation}
where
\begin{equation}\label{3.16}
\alpha=\frac{V}{N}\frac{\varepsilon^{2}(k_{0})\mu\cos2\theta_{c}}{2\pi\hbar^{2}S(k_{0})},
\qquad\cos\theta_{c}=\frac{k_{\textrm{C}}}{2k_{0}}.
\end{equation}
We can see that \emph{the parameter $\rho$ is determined by the
spectrum $E_{ext}(k)$. On the other hand, $\rho$ determines this spectrum.
$E_{ext}(k)$ doesn't contain $\rho$ in an explicit form.
Hence $E_{ext}(k)$ is determined with the parameter $\rho$ by
self-consistent way.}

\begin{figure}[]
\centerline{\includegraphics[width=8cm]{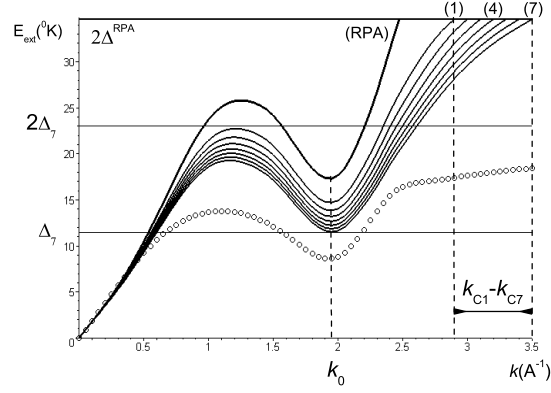}} \caption{The
spectrums of collective excitations. The line (RPA) is spectrum
corresponding to Feynman formula (\ref{2.12}). The lines
\textbf{1}-\textbf{7} are dispersion curves obtained as numerical
solution of the equation (\ref{3.15}) with parameters
$k_{\textrm{C}}$ equaling $2.9A^{-1}$, $3.0A^{-1}$, $3.1A^{-1}$,
$3.2A^{-1}$, $3.3A^{-1}$, $3.4A^{-1}$, $3.5A^{-1}$ accordingly.
The dotted line is experimental spectrum of CE \cite{don} at low
temperature. The dispersion curves \textbf{1}-\textbf{7} finish in the points
$k=k_{\textrm{C}}$ on the line $2\Delta^{RPA}$.}
\label{fig3}
\end{figure}
\begin{figure}[]
\centerline{\includegraphics[width=8cm]{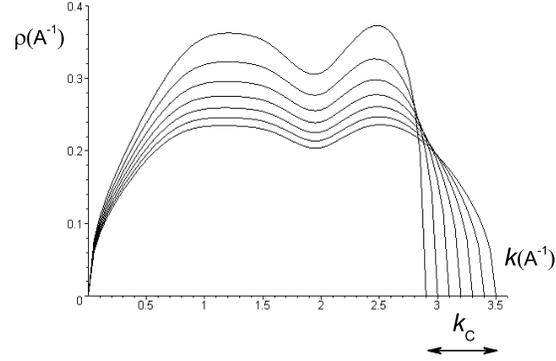}}
\caption{The
parameters of cutt-off $\rho (k)$ for the integral (\ref{3.12}).
The curves $\rho (k)$ correspond to the curves
\textbf{1}-\textbf{7} on Fig.(\ref{fig3}). A domain of the functions
$\rho(k)$ is $0\leq k \leq k_{\textrm{C}}$.}
\label{fig4}
\end{figure}

In the long-wave asymptotic
$E_{ext}^{RPA}(k\rightarrow 0)=c\hbar k+Ak^{3}$, $A=const>0$.
Then it is easily to show that
\begin{eqnarray}\label{3.17}
E_{ext}(k\rightarrow 0)=c\hbar
k-\frac{c\hbar\alpha}{2\chi\Delta^{RPA}}k^{2}+Ak^{3}-\ldots.
\end{eqnarray}
This means that the spectrum $E_{ext}$ is joined with Feynman
formula at small $k$ and describes the sound mode.
Thus the condition (\ref{ii}) is satisfied. However we can see
that even powers of $k$ exist in the expansion (\ref{3.17}). But
this doesn't contradict to isotropy of liquid because the dispersion
equation (\ref{3.15}) depends on modulus of wave vector only
$E_{ext}=E_{ext}(|\textbf{k}|)$.

We can see that the dispersion curve
(\ref{3.15}) depends on the parameter $k_{\textrm{C}}$ (threshold
of decay) strongly. This is \emph{free parameter} being in the limits
$k_{\textrm{C}}^{RPA}\leq k_{\textrm{C}}\leq 2k_{0}$, therefore we have a
family of curves shown in Fig.\ref{fig3}.
Corresponding cut-off parameters $\rho$ which joins the long-wave
asymptotic $k\rightarrow 0$ with the asymptotic $k\rightarrow
k_{\textrm{C}}$ is shown on Fig.\ref{fig4}.

A domain of the function $\rho(k)$ is $0\leq k \leq
k_{\textrm{C}}$ because at $k>k_{\textrm{C}}$ we have
$\rho(k)^{2}<0$. This means that the dispersion curve (\ref{3.15})
can not be continued to the region $k>k_{\textrm{C}}$ (though the
function $E_{ext}(k)$ exists formally). Hence \emph{the point
$k_{\textrm{C}}$ is end point of the spectrum}. For the typical
value $k_{\textrm{C}}=3.2A^{-1}$ we have:
$\max(\rho/k_{0})=0.28/1.93=0.15\ll 1$ and
$\max(\rho/k_{0}/\sin\theta_{c})=0.27<1$. This means that the
condition (\ref{iii}) is satisfied.

The dispersion curves \textbf{1}-\textbf{7} on Fig.\ref{fig3} are more
close to the experimental curve than $E_{ext}^{RPA}(k)$, however they have an
essential defect. CE
with the dispersion curve $E_{ext}(k)$ decays into excitations with spectrum calculated in
RPA: $E_{ext}^{RPA}(q)$ and
$E_{ext}^{RPA}(|\textbf{k}-\textbf{q}|)$. This situation corresponds
to conservation of energy in the form (\ref{3.5}). In reality the
law of conservation of energy is the form (\ref{3.2})
where energy of decaying CE and energies of decay products
are determined with the \emph{real} spectrum. We can see on Fig.\ref{fig3} that the
dispersion curve $E_{ext}(k)$ differs from the curve
$E_{ext}^{RPA}(k)$ essentially in the region of big $k$. So, the
new value of a roton gap is $\Delta<\Delta^{RPA}$. This means that
energy is not conserved at decay of CE into two
rotons. In other words the dispersion curve $E_{ext}(k)$ is not
self-consistent.

\subsection{The self-consistent form of the dispersion curve $E_{ext}(k)$}\label{subSpectrum2}

\begin{figure}[]
\centerline{\includegraphics[width=10.0cm]{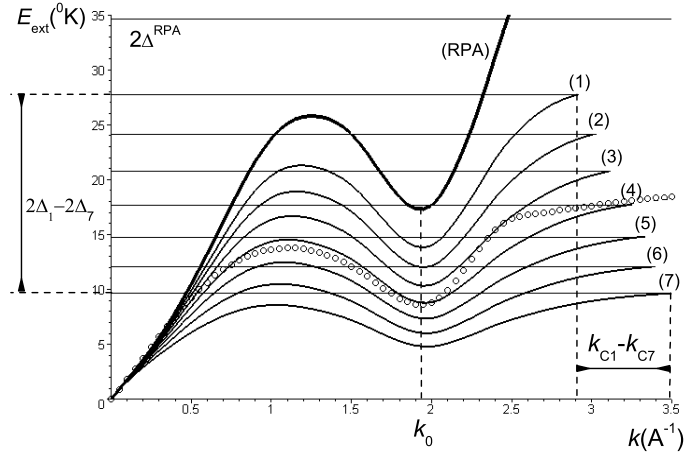}} \caption{The
spectrums of collective excitations. The line (RPA) is spectrum
corresponding to Feynman formula (\ref{2.12}). The lines 1-7 are
dispersion curves obtained by numerical solution of the set of equations
(\ref{3.19}) and (\ref{3.21}) with parameters
$k_{\textrm{C}}=2.9A^{-1}$, $3.0A^{-1}$, $3.1A^{-1}$,
$3.2A^{-1}$, $3.3A^{-1}$, $3.4A^{-1}$, $3.5A^{-1}$ accordingly.
The dotted line is the experimental spectrum of CE \cite{don}.
 The dispersion curves 1-7 are ended
in the points $k=k_{\textrm{C}}$ on the lines
$2\Delta_{1-7}$.} \label{fig5}
\end{figure}

The dispersion equation $E_{ext}(k)$ is self-consistent if
\emph{an energy of the end point of the spectrum is
equal to the doubled roton gap of this spectrum}:
$E_{ext}(k_{\textrm{C}})=2\Delta$. Let the new spectrum has a form
\begin{eqnarray}\label{3.19}
E_{ext}(k)=E_{ext}^{RPA}(k)+\alpha
k\ln\left[\frac{2\widetilde{\Delta}-E_{ext}(k)}{2\widetilde{\Delta}}\right],
\end{eqnarray}
where the parameter $\widetilde{\Delta}$ must be such that the
condition of self-consistency is executed:
\begin{eqnarray}\label{3.20}
\left\{\begin{array}{c}
  E_{ext}(k_{\textrm{C}})=2\Delta \\
  E_{ext}(k_{0})=\Delta \\
\end{array}\right\}.
\end{eqnarray}
The parameter $\Delta$ is the \emph{new} roton gap
in the point $k_{0}$. The the set of equations
(\ref{3.20}) in a expanded form is
\begin{eqnarray}\label{3.21}
\left\{\begin{array}{c}
  2\Delta=E_{ext}^{RPA}(k_{\textrm{C}})+\alpha k_{\textrm{C}}\ln\left(1+\frac{\Delta}{\widetilde{\Delta}}\right) \\
  \Delta=E_{ext}^{RPA}(k_{0})+\alpha
k_{0}\ln\left(1+\frac{\Delta}{2\widetilde{\Delta}}\right)\\
\end{array}\right\}.
\end{eqnarray}
The set of equations (\ref{3.19}) and (\ref{3.21}) determines
a dispersion curve of CE. The results of numerical solution
of these equations with $k_{\textrm{C}}=2.9A^{-1}-3.5A^{-1}$ are
shown on Fig.\ref{fig5}. The curves 1-7 on Fig.\ref{fig5}
correspond to the curves 1-7 on Fig.\ref{fig4}. As in the previous case
the spectrum depends on its end point
$k_{\textrm{C}}$ strongly though sound velocity $\hbar
c=\lim_{k\rightarrow 0}\frac{\partial E_{ext}}{\partial k}$ is the same
in the all cases. We can see a weak pinning of
the curve $E_{ext}(k)$ to the line $2\Delta$. This means
hybridization of the dispersion curve of CE with a two-roton
level \cite{levin}. On the rest of regions an energy $E_{ext}(k)$
is determined by a state with one CE (phonon or roton) completely.
We have the family of curves $E_{ext}$ and we cannot obtain the point
$k_{\textrm{C}}$ on this step of calculations. In order to do this
we must generalize the oscillator model formulated in the
Subsection \ref{subRPA2}.

\section{The ground state and Bose condensate}\label{Vacuum}
\subsection{Formalism of effective mass in the oscillator model.}\label{subVacuum1}

In this section our problem is calculation of a
ground state energy taking into consideration the anharmonic correction
in the equation (\ref{1.13}). If we calculate this
corrections then we have a divergence
$E_{0}=E_{0}^{RPA}+\infty$ as in the Section
\ref{Spectrum}. In order to overcome
this difficulty let's use the oscillator model formulated in the
Section \ref{RPA} where a superfluid liquid
is a totality of independent harmonic oscillators with
frequencies $E_{ext}^{RPA}(k)=\varepsilon(k)/S(k)$. In higher
approximations we have another spectrums of CE $E_{ext}(k)$: the set of equations
(\ref{3.19}) and (\ref{3.21}). However this
spectrum can be written by analogy with RPA as
\begin{equation}\label{4.1}
    E_{ext}(k)=\frac{\tilde{\varepsilon}(k)}{S(k)}, \qquad
\textrm{where}\qquad
\widetilde{\varepsilon}(k)=\frac{\hbar^{2}k^{2}}{2\widetilde{m}(k)}.
\end{equation}
Due to contribution of anharmonic terms of the Hamiltonian
(\ref{1.9}) mass of a \emph{particle} $m$ is renormalized to mass
of a \emph{quasi-particle} $\widetilde{m}(k)$.
Continuing the analogy with RPA we can write the effective
Hamiltonian having a form as in RPA:
\begin{eqnarray}\label{4.2}
&&\hat{H}_{\textrm{eff}}=\sum_{\textbf{k}_{1}\neq
0}\widetilde{\varepsilon}(k_{1})\left(
-\frac{\partial^{2}}{\partial\rho_{\textbf{k}_{1}}\partial\rho_{-\textbf{k}_{1}}}+\frac{1}{4}\rho_{\textbf{k}_{1}}\rho_{-\textbf{k}_{1}}-\frac{1}{2}\right)
\nonumber\\
&&+\frac{N^{2}}{2V}\widetilde{\nu}(0)+\frac{N}{2V}\sum_{\textbf{k}_{1}\neq
0}\widetilde{\nu}(k_{1})\left(\rho_{\textbf{k}_{1}}\rho_{-\textbf{k}_{1}}-1\right)
\end{eqnarray}
and the corresponding Jacobian:
\begin{equation}\label{4.3}
\ln J=\ln\left(
V^{N}\prod'_{\textbf{k}\neq0}\frac{1}{\pi}\right)-\frac{1}{2}\sum_{\textbf{k}\neq
0}\rho_{\textbf{k}}\rho_{-\textbf{k}},
\end{equation}
where the normalization requirement (\ref{1.5}) has been used.

\begin{figure}[]
\centerline{\includegraphics[width=8cm]{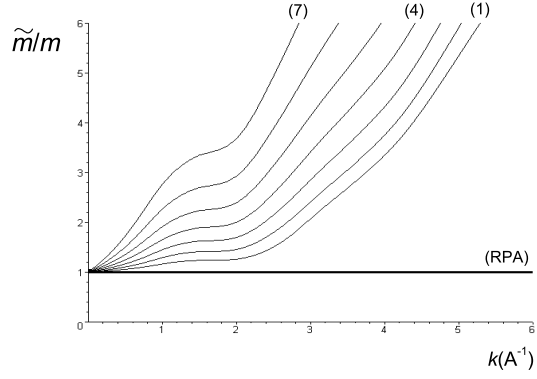}} \caption{The
effective masses of quasi-particles in units of mass of $^{4}He$:
$\widetilde{m}/m$. The curve (RPA) corresponds to a "naked"
particle. The numbers 1-7 correspond to curves in Fig.\ref{fig5}.
The functions $\widetilde{m}(k)$ have a sense in the range
$k<k_{\textrm{C}}$ only.} \label{fig6}
\end{figure}

\begin{figure}[]
\centerline{\includegraphics[width=8cm]{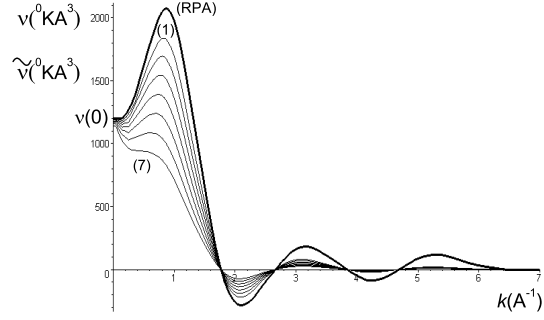}} \caption{The
potential of interaction between particles (RPA) and between
quasi-particles (1-7).} \label{fig7}
\end{figure}

\emph{We must
obtain the same structure factor $S(k)$ in any
approximation because this function is given by experiment}.
Therefore we must introduce the effective interaction $\tilde{\nu}(k)$
as follows:
\begin{eqnarray}\label{4.4}
&&S(k)=-\frac{1}{4f(k)}=\frac{1}{\sqrt{1+\frac{2N}{V}\frac{\nu(k)}{\varepsilon(k)}}}
=\frac{1}{\sqrt{1+\frac{2N}{V}\frac{\widetilde{\nu}(k)}{\widetilde{\varepsilon}(k)}}}\nonumber\\
&&\Rightarrow\widetilde{\nu}(k)=\widetilde{\varepsilon}(k)\frac{V}{2N}\left[\frac{1}{S(k)^{2}}-1\right].
\end{eqnarray}
The effective masses and interactions are shown in Fig.\ref{fig6}
and Fig.\ref{fig7}. It is necessary to notice that
$\tilde{m}(k\rightarrow 0)=m$ and $\tilde{\nu}(0)=\nu(0)$.


As a result we have the totality of harmonic oscillators again.
These oscillators have the same dispersions
$\langle\rho_{\textbf{k}}\rho_{-\textbf{k}}\rangle$ as in RPA.
However they have another frequencies $E_{ext}(k)$ instead of the
old $E_{ext}^{RPA}(k)$. Difference between new frequencies and the
initial frequencies depends on the aforesaid anharmonicities. A motion equation for each oscillator can be written as the
equation (\ref{2.13}) but with an effective mass and an effective
interaction:
\begin{eqnarray}\label{4.5}
-\frac{\hbar^{2}}{2\widetilde{M}(k)}\frac{\partial^{2}\bar{\psi}}{\partial\rho_{\textbf{k}}\partial\rho_{-\textbf{k}}}
+\left[\frac{1}{4}\widetilde{\varepsilon}(k)+\frac{N}{2V}\widetilde{\nu}(k)\right]\rho_{\textbf{k}}\rho_{-\textbf{k}}\bar{\psi}=\widetilde{\xi}(k)\bar{\psi}.
\end{eqnarray}
A ground state energy can be rewritten as follows:
\begin{eqnarray}\label{4.6}
E_{0}=\sum_{\textbf{k}\neq
0}\frac{1}{2}\frac{\widetilde{\varepsilon}(k)}{S(k)}
-\sum_{\textbf{k}\neq
0}\left[\frac{1}{2}\widetilde{\varepsilon}(k)+\frac{N}{2V}\widetilde{\nu}(k)\right]+\frac{N^{2}}{2V}\widetilde{\nu}(0).
\end{eqnarray}
It is necessary to notice that we must do transition from
summation on $\textbf{k}$ to integration as follows:
\begin{equation}\label{4.6b}
\sum_{\textbf{k}\neq
0}\rightarrow\frac{V}{(2\pi)^{3}}\int_{0}^{k_{\textrm{C}}}k^{2}dk.
\end{equation}
The results of calculation of a ground state energy (per one
atom) for particles' system (RPA) and for quasi-particles' system
(the curves \textbf{1}-\textbf{7}) are:
\begin{eqnarray}\label{4.7}
&&E_{0}^{RPA}/N=-13.78 ^{o}K, \qquad E_{0}^{(1)}/N=-8.73^{o}K, \qquad E_{0}^{(2)}/N=-6.66^{o}K \nonumber\\
&&E_{0}^{(3)}/N=-4.62^{o}K \qquad E_{0}^{(4)}/N=-2.60^{o}K \qquad E_{0}^{(5)}/N=-0.59^{o}K \nonumber\\
&&E_{0}^{(6)}/N=1.43^{o}K \qquad E_{0}^{(7)}/N=3.43^{o}K
\end{eqnarray}
We can see that existence of the end point of a dispersion curve
causes some increase of a ground state energy. Since
$k_{\textrm{C}}$ is unknown then we cannot select a value of
this energy. The energy $E_{0}$ is
minimal at $k_{\textrm{C}}<2.9A^{-1}$. But the important condition for
oscillator frequencies $\hbar\omega=E_{ext}$ exists yet.

\begin{figure}[]
\centerline{\includegraphics[width=10cm]{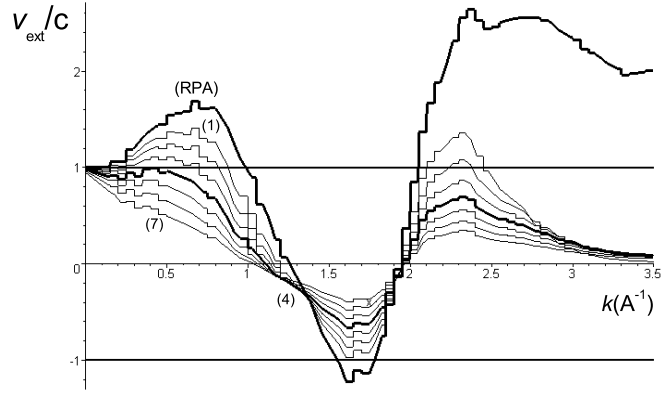}} \caption{The
velocities of collective excitations $v_{ext}(k)=\frac{\partial
E_{ext}(k)}{\hbar\partial k}$ per units of a sound velocity $c$.
$v_{ext}(k)<0$ means that the velocity and the momentum
$\textbf{k}$ have contrary directions. The curve (RPA) corresponds
to Feynman spectrum. The curves
\textbf{1}-\textbf{7} correspond to the curves on Fig.\ref{fig5}.
The condition (\ref{4.7a}) is satisfied for the curves where
$|v_{ext}(k)|/c<1$ for all $k$. The spectrums
\textbf{4}-\textbf{7} are stable about radiation of a phonon. The
spectrum \textbf{4} (with $k_{\textrm{C}}=3.2 A^{-1}$, it is
marked by bold line) is on the limit of stability.} \label{fig8}
\end{figure}

If velocity of a CE $\frac{\partial E_{ext}}{\partial
p}\equiv\frac{\partial \hbar\omega}{\partial p}$ is more than a
sound velocity in a system then CE radiates
a phonon \cite{lif}. The regions of a dispersion curve, where
such is possible, is unstable. We have formulated the
oscillator model of Bose liquid where its ground state energy
is a sum of energies of ground states of oscillators representing
the liquid. In other words the ground state is an infinity number
of \emph{virtual} collective excitations. Then \emph{in order to
determine a ground state as a state with minimal energy a
dispersion curve $E_{ext}(k)$ must be stable}.
It means that for all $k$ the inequality must be executed:
\begin{equation}\label{4.7a}
\frac{\partial E_{ext}(k)}{\partial k}\equiv<\lim_{k\rightarrow 0}\frac{\partial
E_{ext}(k)}{\partial k}=c\hbar,
\end{equation}
If this condition is not satisfied then the ground
state falls apart. The plots of CE's velocity
is shown in Fig.\ref{fig8}. The curves \textbf{4}-\textbf{7} satisfy the criterium
(\ref{4.7a}). Hence possible end points present in the interval
$k_{\textrm{C}}=3.2A^{-1}\div 2k_{0}$ only. The spectrum
\textbf{4} corresponds to the minimal value $k_{\textrm{C}}=3.2A^{-1}$.
On the contrary, the curve (RPA) is unstable almost
completely.

\begin{figure}[]
\centerline{\includegraphics[width=10cm]{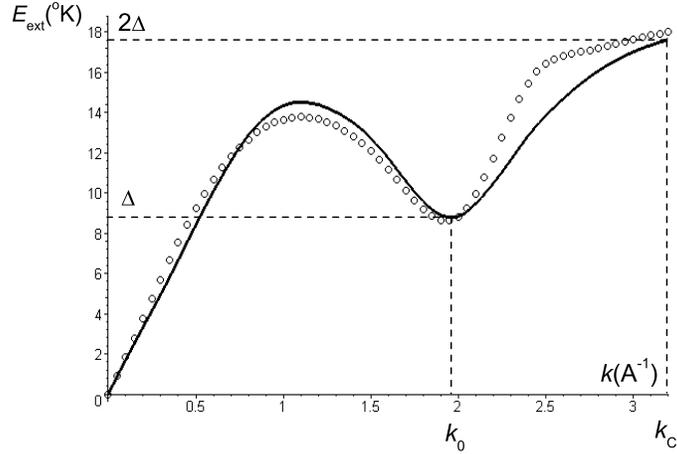}} \caption{The
dispersion curve of CE obtained
with the optimal parameter $k_{\textrm{C}}=3.2A^{-1}$. As result
we have: $\mu=0.19m$ (the new roton mass), $\Delta=8.8^{o}K,
k_{0}=1.95A^{-1}$. The dotted line is experimental spectrum of CE
\cite{don} at low temperature. In the point
$k=k_{\textrm{C}}$ on the line $2\Delta$ the dispersion curve
ends.}\label{fig9}
\end{figure}

We can see in (\ref{4.7}) that increase of $k_{\textrm{C}}$ causes
increase of $E_{0}$:
$E_{0}^{(4)}<E_{0}^{(5)}<E_{0}^{(6)}<E_{0}^{(7)}$. Hence \emph{the
energy $E_{0}^{(4)}$ is minimal for all allowed energies}
$E_{0}$. This means that
\begin{eqnarray}\label{4.7b}
E_{0}/N=E_{0}^{(4)}/N=-2.60^{o}K.
\end{eqnarray}
The experimental value is $E_{0}/N=-7.12^{o}K$. It is necessary to notice
that short-range correlations make essential contributions in $E_{0}$. These
correlations are calculated badly by the collective
variables method. So, in \cite{vacar6} it has been purposed the hybrid approach
where transition to collective
variables was made not completely and part of the coordinates
was kept as Cartesian coordinates. Proceeding from aforesaid
we can consider the result (\ref{4.7b}) as satisfactory.

The final spectrum of a collective excitation $E_{ext}(k)$ is
shown in Fig.\ref{fig9}. The dispersion curve has been obtained as
a numerical solution of the set of equations (\ref{3.19})
and (\ref{3.21}) with the optimal parameter
$k_{\textrm{C}}=3.2A^{-1}$. We can see a weak
pinning of the dispersion curve to the line $2\Delta$ that
means decay of CE into
two rotons with energy $\Delta$ and momentum $k_{0}$ in
each. The dispersion curve coincides with the experimental curve
well.

\subsection{Suppression of Bose condensate.}\label{subVacuum2}

In Bose system at sufficiently low
temperature the occupation number $N_{0}$ of a state with minimal
energy (momentum of particles is $\mathbf{p}=0$)
is macroscopic number $N_{0}\leq N$.
For ideal gas density of a condensate and temperature of the
transition are
\begin{equation}\label{4.8}
\frac{N_{0}}{N}=1-\left(\frac{T}{T_{0}}\right)^{3/2},\qquad T_{0}=\frac{%
2\pi\hbar^{2}}{m}\left(\frac{N/V}{2.612}\right)^{2/3}.
\end{equation}
At temperature $T=0$ all atoms are in BC: $N_{0}=N$.
If particles interact then BC is suppressed. BC
is suppressed strongly in HeII: $N_{0}/N=0.07$. In the paper \cite{pash3}
the model was proposed where BC is suppressed due to formation of pairs of boson.
In this section we shall consider suppression of BC from the standpoint of collective motions of
liquid using the oscillator model.

The equations (\ref{2.13}) or
(\ref{4.5}) describe harmonic oscillations of Bose liquid.
According to the uncertainty principle the energy of a oscillator
in a ground state can't be zero $1/2\hbar\omega\neq 0$. The energy
is divisible by kinetic energy and potential energy in equal parts
$\langle T\rangle=\langle U\rangle =M\omega^{2}\langle
\rho_{\textbf{k}}\rho_{-\textbf{k}}\rangle /2=\xi/2$ according to
the virial theorem. This zero point energy is the energy of
fluctuations of a quantum liquid. Hence collective
motions is in a ground state and no all particles can be in
BC. That's BC is suppressed by the dynamical fluctuations
\cite{grig}.

Using the equation (\ref{2.3}) let's write the energy of a ground
state in the a form:
\begin{eqnarray}\label{4.9}
&&\sum_{\textbf{k}\neq
0}\varepsilon(k)\left[-4f^{2}(k)\rho_{\textbf{k}}\rho_{-\textbf{k}}-2f(k)
+\frac{1}{4}\rho_{\textbf{k}}\rho_{-\textbf{k}}-\frac{1}{2}\right]\nonumber\\
\nonumber\\
&&+\frac{N^{2}}{2V}\nu_{0}+\sum_{\textbf{k}\neq
0}\frac{N}{2V}\nu_{k}(\rho_{\textbf{k}}\rho_{-\textbf{k}}-1)=E_{0}.
\end{eqnarray}
After ground state averaging of this expression according to the
rule
$\rho_{\textbf{k}}\rho_{-\textbf{k}}\rightarrow\langle\rho_{\textbf{k}}\rho_{-\textbf{k}}\rangle$,
taking into consideration the expression for the potential energy
of a system
\begin{equation}\label{4.10}
\sum_{1\leq i<j\leq
N}\Phi(|\textbf{r}_{i}-\textbf{r}_{j}|)=\frac{N^{2}}{2V}\nu_{0}+\sum_{\textbf{k}\neq
0}\frac{N}{2V}\nu_{k}(\langle\rho_{\textbf{k}}\rho_{-\textbf{k}}\rangle-1)
\end{equation}
and the evident equality $E_{0}=\langle
T\rangle+\langle\Phi\rangle$, we can write the average kinetic
energy of a system as
\begin{equation}\label{4.11}
\langle T\rangle=\sum_{\textbf{k}\neq
0}\varepsilon(k)\left[\left(-4f^{2}(k)+\frac{1}{4}\right)\langle\rho_{\textbf{k}}\rho_{-\textbf{k}}\rangle-2f(k)
-\frac{1}{2}\right].
\end{equation}
Since $\varepsilon(k)$ is kinetic energy of a particle then the
expression in the square brackets
can be understood as occupation numbers of above-condensate
particles $N_{\textbf{k}\neq0}$. Using the connection
(\ref{2.7}) we can write the kinetic energy and the corresponding
occupation number via the function $f(k)$ or via the structure
factor $S(k)$:
\begin{eqnarray}
\langle
T\rangle=\sum_{\textbf{k}\neq0}\frac{\hbar^{2}k^{2}}{2m}\left[\frac{(4f(k)+1)^{2}}{-16f(k)}\right],\label{4.12a}\\
\nonumber\\
N_{\textbf{k}\neq0}=\frac{(4f(k)+1)^{2}}{-16f(k)}=\frac{(S(k)-1)^{2}}{4S(k)}.\label{4.12b}
\end{eqnarray}
Then the density of BC is
\begin{equation}\label{4.13}
N_{0}=N-\sum_{\textbf{k}\neq0}N_{\textbf{k}}=N-\frac{V}{(2\pi)^{3}}\int_{0}^{\infty}\frac{(S-1)^{2}}{4S}4\pi
k^{2}dk.
\end{equation}
In the case of non-interacting particles we have: $f(k)=-1/4$ and
$S(k)=1$, hence $N_{\textbf{k}}=0$, $N_{0}=N$. In the case of weakly
non-ideal gas with the interaction
$\nu_{\textbf{k}}=\frac{4\pi\hbar^{2}}{m}a$ if a scattering length
$\frac{a^{3}}{V/N}\ll 1$ the density of BC is
\begin{equation}\label{4.14}
    \frac{N_{0}}{N}=1-\frac{8}{3}\sqrt{\frac{Na^{3}}{\pi V}}
\end{equation}
that coincides with the result of Bogoliubov \cite{lif}.

The expression for density of BC (\ref{4.13}) is correct for
systems with weak interaction only. As it has been shown in
\cite{vacar4,vacar5} for a total case we must use the expression
\begin{equation}\label{4.16}
    \frac{N_{0}}{N}=F(R\rightarrow\infty)=\exp\left(-\frac{V}{(2\pi)^{3}N}\int_{0}^{\infty}N_{\textbf{k}}4\pi
k^{2}dk\right),
\end{equation}
where $F(\textbf{r}|\textbf{r}')$ is a one-particle density
matrix, $R=|\textbf{r}'-\textbf{r}|$.
The expression (\ref{4.13}) is the two first terms of an expansion
of the exponent (\ref{4.16}). In RPA the density
of BC is
\begin{equation}\label{4.18}
\left(\frac{N_{0}}{N}\right)_{RPA}=0.263.
\end{equation}
In the higher approximations we have the effective Hamiltonian
(\ref{4.2}). In order to calculate density of BC with the
effective Hamiltonian $\widehat{H}_{\textrm{eff}}$ (\ref{4.2}) we
are going to proceed from the following reason. Let's write the
energy of a ground state of Bose system in a form:
\begin{eqnarray}\label{4.20}
\sum_{\textbf{k}\neq0}\widetilde{\varepsilon}(k)\left[\frac{(S(k)-1)^{2}}{4S(k)}\right]+\frac{N^{2}}{2V}\nu_{0}+\sum_{\textbf{k}\neq
0}\frac{N}{2V}\widetilde{\nu}_{k}[S(k)-1]=E_{0}.
\end{eqnarray}
We must mark out the kinetics energy of a system of
\emph{particles}. All anharmonic terms in the Hamiltonian
$\widehat{H}$(\ref{1.9}) or in the Hamiltonian
$\widehat{H}_{\textrm{BZ}}$ (\ref{1.7a}) are the terms of a
operator of kinetic energy $\sum_{1\leq j\leq
N}\frac{\hat{p}^{2}_{j}}{2m}$. This means that renormalization of
mass $m\rightarrow \widetilde{m}$ and interaction $\nu(k)
\rightarrow\widetilde{\nu}(k)$ occurs due to the contribution of
kinetic energy but not the interaction as in usual perturbation
theory. Hence the potential energy of a system $\sum_{1\leq
i<j\leq N}\Phi(|\textbf{r}_{i}-\textbf{r}_{j}|)$ (\ref{4.10})
does't change at transition from $\widehat{H}$ to
$\widehat{H}_{\textrm{eff}}$. In order to obtain the new kinetic
energy of particles let's rewrite (\ref{4.20}) in the identical
form:
\begin{eqnarray}\label{4.21}
&&\sum_{\textbf{k}\neq0}\left(\widetilde{\varepsilon}(k)\left[\frac{(S(k)-1)^{2}}{4S(k)}\right]+\frac{N}{2V}(\widetilde{\nu}_{k}-\nu_{k})[S(k)-1]\right)\nonumber\\
&&+\frac{N^{2}}{2V}\nu_{0}+\sum_{\textbf{k}\neq
0}\frac{N}{2V}\nu_{k}[S(k)-1]=E_{0}.
\end{eqnarray}
Proceeding from aforesaid the first term is a new kinetic energy
of particles. In order to obtain a momentum distribution $N_{\textbf{k}}$
of particles let's mark out kinetic energy of a
\emph{particle} $\varepsilon(k)$ in this expression. Then we have
\begin{eqnarray}
&&\langle T\rangle=\label{4.22a}
\sum_{\textbf{k}\neq0}\varepsilon(k)\left(\frac{\widetilde{\varepsilon}(k)}{\varepsilon(k)}\left[\frac{(S(k)-1)^{2}}{4S(k)}\right]
+\frac{N}{2V}\frac{\widetilde{\nu}_{k}-\nu_{k}}{\varepsilon(k)}[S(k)-1]\right),
\end{eqnarray}
\begin{eqnarray}
&&N_{\textbf{k}\neq0}=\label{4.22b}
\frac{\widetilde{\varepsilon}(k)}{\varepsilon(k)}\left[\frac{(S(k)-1)^{2}}{4S(k)}\right]
+\frac{N}{2V}\frac{\widetilde{\nu}_{k}-\nu_{k}}{\varepsilon(k)}[S(k)-1].
\end{eqnarray}

Substituting (\ref{4.22b}) in (\ref{4.16}) we obtain
the densities of BC for the dispersion curves
\textbf{1}-\textbf{7} in Fig.\ref{fig5}:
\begin{eqnarray}\label{4.24}
&&N_{0}^{(1)}/N=0.149, \qquad N_{0}^{(2)}/N=0.104, \qquad N_{0}^{(3)}/N=0.071\nonumber\\
&&\textbf{N}_{\textbf{0}}^{\textbf{(4)}}\textbf{/N}\textbf{=0.048}\\
&&N_{0}^{(5)}/N=0.031 \qquad N_{0}^{(6)}/N=0.020 \qquad
N_{0}^{(7)}/N=0.013\nonumber
\end{eqnarray}

As it has been said before, the curve \textbf{4} with
$k_{\textrm{C}}=3.2A^{-1}$ realizes in practice only. This means
that portion of BC in superfluid helium at zeroth temperature is
$N_{0}/N=0.048$. Density of BC measured in experiment by the
method of deep inelastic scattering of neutrons \cite{gly} and by
the method of quantum evaporation \cite{wyatt} is $0.07$.
Hence the result (\ref{4.24}) is satisfactory fully.

\section{Conclusion}

As a result of self-consistent solution of Schrodinger equation
for the state of Bose liquid with one CE the dispersion law $E_{ext}(k)$ (the set of equations
(\ref{3.19}) and (\ref{3.21})) has been obtained.  It has
been shown that for the Hermitian form
of Bogoliubov-Zubarev Hamiltonian the second term in the Bijl-Dingle-Jastrow
expansion considers real
and virtual processes of decay of CE. Higher
terms in this expansion considers decays into three and more excitations that is very
improbable processes. Decay of CE on two rotons causes existence of the end
point $k_{\text{C}}$ of a dispersion curve. Our main result is: \emph{the
phonon-roton dispersion curve $E_{ext}(k)$ is determined by both
interaction between bosons $\nu(k)$ and the end point}
$k_{\textrm{C}}$. \emph{That is the dispersion curve strongly
depends on property of its stability}.

In order to find $k_{\text{C}}$ and the energy of ground state we
have been formulated the oscillator model of Bose liquid.
According to this model the basic B-Z Hamiltonian can by rewritten as
Hamiltonian in the harmonic approximation with renormalized kinetic and potential energies.
As a result we have a totality of
harmonic oscillators again with the same dispersions $S(k)$ but with
another frequencies already. Energy of a ground state $E_{0}$
depends on these frequencies, hence it depends on the end point
$k_{\textrm{C}}$. Proceeding from the condition of minimum of
energy and from the condition of stability of
ground state we can obtain the unknown end point
$k_{\textrm{C}}=3.2\textrm{A}^{-1}$ of the spectrum.
The obtained spectrum of CE $E_{ext}$ is a
function connecting energy of a excitation with the structure factor
$S(k)$ of liquid which is taken from the experiment. Thus we did not
use any model potentials of interaction. It means that all our
calculations have been done without any adaption parameters.

A mechanism of suppression of one-particle BC has been described
at temperature of absolute zero. As a consequence of the
uncertainty principal the dynamical fluctuations exist in a
ground state and give nonzero kinetic energy. These fluctuations
suppresses BC. It has been shown that such fluctuations are absent
for ideal gas. It is necessary to notice that the analogy
dynamical quantum fluctuations in other systems can cause quantum
phase transitions \cite{stish}. At nonzero
temperatures BC is suppressed by both dynamical fluctuations
and kinematic (thermal) fluctuations.

\appendix
\section{The operator of momentum}\label{A}

Let's consider the expression:
%\begin{widetext}
\begin{eqnarray}\label{A1}
&&
J^{1/2}\hat{\textbf{P}}J^{-1/2}\bar{\psi}=-\sum_{\textbf{k}_{1}\neq
0}\hbar\textbf{k}_{1}\rho_{\textbf{k}_{1}}J^{1/2}\frac{\partial}{\partial\rho_{\textbf{k}_{1}}}\left[J^{-1/2}\bar{\psi}\right]
=\frac{1}{2}\sum_{\textbf{k}_{1}\neq
    0}\hbar\textbf{k}_{1}\rho_{\textbf{k}_{1}}\frac{\partial\ln
    J}{\partial\rho_{\textbf{k}_{1}}}+
    \hat{\textbf{P}}\bar{\psi}\nonumber\\
    &&=\frac{1}{2}\sum_{\textbf{k}\neq 0}\sum_{\textbf{k}_{1}\neq
    0}\hbar\textbf{k}_{1}
    \rho_{\textbf{k}_{1}}\frac{\partial}{\partial\rho_{\textbf{k}_{1}}}\left[\ln
    C-\frac{1}{2}\rho_{\textbf{k}}\rho_{-\textbf{k}}\right]+\frac{1}{2}\sum_{\textbf{k}\neq
    0}\sum_{\textbf{q}\neq 0}\sum_{\textbf{k}_{1}\neq
    0}\hbar\textbf{k}_{1}
    \rho_{\textbf{k}_{1}}\frac{\partial}{\partial\rho_{\textbf{k}_{1}}}\left[\frac{1}{6}
    \rho_{\textbf{k}}\rho_{\textbf{q}-\textbf{k}}\rho_{-\textbf{q}}\right]\nonumber\\
    &&+\ldots +
    \hat{\textbf{P}}\bar{\psi}=-\frac{1}{2}\sum_{\textbf{k}\neq
0}[\hbar\textbf{k}\rho_{\textbf{k}}\rho_{-\textbf{k}}
-\hbar\textbf{k}\rho_{\textbf{k}}\rho_{-\textbf{k}}]+\frac{1}{6}\sum_{\textbf{k}\neq
    0}\sum_{\textbf{q}\neq
0}[\hbar\textbf{k}\rho_{\textbf{k}}\rho_{\textbf{q}-\textbf{k}}\rho_{-\textbf{q}}\nonumber\\
&&+\hbar(\textbf{q}-\textbf{k})
\rho_{\textbf{k}}\rho_{\textbf{q}-\textbf{k}}\rho_{-\textbf{q}}+
\hbar(-\textbf{q})\rho_{\textbf{k}}\rho_{\textbf{q}-\textbf{k}}\rho_{-\textbf{q}}]
-\ldots+\hat{\textbf{P}}\bar{\psi}=0+\hat{\textbf{P}}\bar{\psi}\Longrightarrow
\hat{\textbf{P}}\bar{\psi}=\textbf{P}\bar{\psi}.
\end{eqnarray}
%\end{widetext}
That is if the function $\psi$ is eigenfunction of the operator of
momentum $\hat{\textbf{P}}$ then $\bar{\psi}$ is eigenfunction of
the operator of momentum too.

\section{The wave function of a system of free bosons}\label{B}

If to exclude interaction between particles: $\nu(k)=0$, the wave
function of a system $\psi=\bar{\psi}J^{-1/2}$ must be a wave
function of free bosons. So, we can write for a ground state:
\begin{equation}\label{B1}
    \psi=Be^{U}J^{-1/2}=const\Rightarrow U=\frac{1}{2}\ln J,
\end{equation}
where $B$ is a normalization constant. Let's verify it. In a
ground state of a system of noninteracting bosons $E_{0}=0$. Then
proceeding from the equation (\ref{1.13}) we can write:
\begin{eqnarray}\label{B2}
&&-\frac{\partial^{2}U}{\partial\rho_{\textbf{k}_{1}}\partial\rho_{-\textbf{k}_{1}}}-\frac{\partial
U}{\partial\rho_{\textbf{k}_{1}}}\frac{\partial
U}{\partial\rho_{-\textbf{k}_{1}}}-\frac{1}{4}\rho_{\textbf{k}_{1}}\frac{\partial
\ln J}{\partial\rho_{\textbf{k}_{1}}}-\frac{1}{2}+
\nonumber \\
\\
&&\sum_{\textbf{k}_{2}\neq
0}\frac{\textbf{k}_{1}\cdot\textbf{k}_{2}}{k^{2}_{1}\sqrt{N}}\rho_{\textbf{k}_{1}
+\textbf{k}_{2}}\left[\frac{\partial
U}{\partial\rho_{\textbf{k}_{1}}\partial\rho_{\textbf{k}_{2}}}+\frac{\partial
U}{\partial\rho_{\textbf{k}_{1}}}\frac{\partial
U}{\partial\rho_{\textbf{k}_{2}}}\right]=0\nonumber.
\end{eqnarray}
Now let's rewrite the third term of the equation (\ref{B2}) in the
form:
\begin{eqnarray}\label{B3}
&&-\frac{1}{4}\frac{\partial \ln
J}{\partial\rho_{\textbf{k}_{1}}}\left[-\frac{\partial \ln
J}{\partial\rho_{-\textbf{k}_{1}}}+\frac{1}{\sqrt{N}}\sum_{\textbf{k}\neq
0}\frac{\textbf{k}\textbf{k}'}{k^{2}}\rho_{\textbf{k}+\textbf{k}'}\frac{\partial
\ln J}{\partial\rho_{\textbf{k}'}}\right]
\nonumber \\
\\
&&=\frac{\partial U}{\partial\rho_{\textbf{k}_{1}}}\frac{\partial
U}{\partial\rho_{-\textbf{k}_{1}}}-\sum_{\textbf{k}_{2}\neq
0}\frac{\textbf{k}_{1}\cdot\textbf{k}_{2}}{k^{2}_{1}\sqrt{N}}\rho_{\textbf{k}_{1}
+\textbf{k}_{2}}\frac{\partial
U}{\partial\rho_{\textbf{k}_{1}}}\frac{\partial
U}{\partial\rho_{\textbf{k}_{2}}},\nonumber
\end{eqnarray}
where we used the equation for the Jacobian (\ref{1.10}) and the
equality (\ref{B1}). After that we have the equation:
\begin{equation}\label{B4}
-\frac{\partial^{2}U}{\partial\rho_{\textbf{k}_{1}}\partial\rho_{-\textbf{k}_{1}}}-\frac{1}{2}+
\sum_{\textbf{k}_{2}\neq
0}\frac{\textbf{k}_{1}\cdot\textbf{k}_{2}}{k^{2}_{1}\sqrt{N}}\rho_{\textbf{k}_{1}
+\textbf{k}_{2}}\frac{\partial
U}{\partial\rho_{\textbf{k}_{1}}\partial\rho_{\textbf{k}_{2}}}=0.
\end{equation}
With help the relationship (\ref{B1}) we can write:
\begin{equation}\label{B5}
-\frac{1}{2}\frac{\partial}{\partial\rho_{\textbf{k}_{1}}}\left[
\rho_{\textbf{k}_{1}}+\frac{\partial\ln
J}{\partial\rho_{-\textbf{k}_{1}}}-\frac{1}{\sqrt{N}}\sum_{\textbf{k}_{1}\neq
0}\frac{\textbf{k}_{1}\textbf{k}_{2}}{k^{2}_{1}}\rho_{\textbf{k}_{1}+\textbf{k}_{2}}\frac{\partial
\ln J}{\partial\rho_{\textbf{k}_{2}}}\right]=0,
\end{equation}
because the expression in the square brackets coincides with the
equation (\ref{1.10}). We can see that the wave function
(\ref{B1}) satisfies to the Schrodinger equation (\ref{1.13}).

Let's obtain normalization constant $B$. Let the wave function of
a ground state $\bar{\psi}$ (\ref{2.2}) is determined by the
following way in RPA:
\begin{equation}\label{B6}
\bar{\psi}=A\exp\left[-\frac{1}{4}\sum_{\textbf{k}\neq0}\frac{1}{S(k)}\rho_{\textbf{k}}\rho_{-\textbf{k}}\right],
\end{equation}
where (\ref{2.7}) was taken into account. The constant $A$ is
obtained from the normalization condition (\ref{1.7}) in the
thermodynamical limit $N\rightarrow\infty$:
\begin{eqnarray}\label{B7}
\prod'_{\textbf{k}\neq
0}\int_{-\infty}^{\infty}d\rho_{\textbf{k}}^{c}\int_{-\infty}^{\infty}d\rho_{\textbf{k}}^{c}|\bar{\psi}|^{2}=1
\Rightarrow A=\prod_{\textbf{k}\neq0}'\sqrt{\frac{1}{\pi S(k)}}.
\end{eqnarray}
Let's write the Jacobian in RPA
\begin{equation}\label{B10}
J=C\exp\left[-2\sum_{\textbf{k}\neq0}'(\rho_{\textbf{k}}^{ё})^{2}+(\rho_{\textbf{k}}^{s})^{2}\right].
\end{equation}
The constant $C$ is obtained from the condition (\ref{1.5}) in the
thermodynamical limit $N\rightarrow\infty$:
\begin{equation}\label{B11}
    V^{N}=\prod'_{\textbf{k}\neq
    0}\int_{-\infty}^{\infty}d\rho_{\textbf{k}}^{c}\int_{-\infty}^{\infty}d\rho_{\textbf{k}}^{c}J
    \Rightarrow C=V^{N}\prod_{\textbf{k}\neq0}'\frac{1}{\pi}.
\end{equation}
Then
\begin{eqnarray}\label{B.13}
\psi=J^{-1/2}\bar{\psi}=\frac{1}{\sqrt{V^{N}}}\prod_{\textbf{k}\neq0}\frac{1}{\sqrt{S(k)}}
\exp\left(-\frac{1}{4}\sum_{\textbf{k}\neq0}\left[\frac{1}{S(k)}-1\right]\rho_{\textbf{k}}\rho_{-\textbf{k}}\right).
\end{eqnarray}
If interaction between particles is absent, then $S(k)=1$ and the
wave function of a system $\psi$ transforms to the wave function
of a system of free bosons $\psi_{\nu=0}=1/\sqrt{V^{N}}$, and the
normalization constant is
\begin{equation}\label{B14}
B=1/\sqrt{V^{N}}.
\end{equation}
The constant $B$ is the same for higher powers of expansion of the
wave function and Jacobian.

\end{document}